\documentstyle [12pt,epsf] {article}
\newcommand {\nc} {\newcommand}

 \newcommand\beq{\begin{equation}}
 \newcommand\eeq{\end{equation}}

 \def\beqn{\begin{eqnarray}}
 \def\eeqn{\end{eqnarray}}

\nc{\re} {\mathop{\mathrm{Re}}} \nc{\im} {\mathop{\mathrm{Im}}} 
\nc{\case}[2] {\mbox{$\frac{#1}{#2}$}} 

  \nc {\gap} {\\[1ex]}
  
  \nc {\f} {\frac}                \nc {\s} {\sqrt}
  
  \nc {\kap} {\kappa}		
  \nc {\amp}[1] {\phi_{#1}}
  \nc {\sgmtot} {\sigma_{\rm tot}}
  \nc {\aml}[2] {\phi_{#1}^{\mathrm{#2}}}
  \nc {\text}[1] {\mbox{\small{#1}}}		
  \nc {\delsigL} {\Delta\sigma_{_{\mathrm L}}}  
  \nc {\delsigT} {\Delta\sigma_{_{\mathrm T}}}  

\begin{document}
\bibliographystyle{unsrt}

\pagestyle{empty}               
	
\rightline{\vbox{
	\halign{&#\hfil\cr
	&RHIC Spin Note\cr}}}

\rightline{\vbox{
	\halign{&#\hfil\cr
	&September 27, 2005\cr}}}
\vskip 1in
\begin{center}
{\Large\bf
{Double-spin asymmetry in elastic proton-proton scattering as a probe for the Odderon}}
\vskip .5in
\normalsize
T.L.\ Trueman \footnote{This manuscript has been authored
under contract number DE-AC02-76CH00016 with the U.S. Department
of Energy.  Accordingly, the
U.S. Government retains a non-exclusive, royalty-free license to
publish or reproduce the published form of this contribution, or
allow others to do so, for U.S. Government purposes.}\\
{\sl Physics Department, Brookhaven National 
Laboratory, Upton, NY 11973}
\end{center}
\vskip 0.5
in
\begin{abstract} The magnitude of the double-spin asymmetry $A_{NN}$  in polarized proton elastic scattering is estimated using results from an earlier determination of the spin-flip coupling of the leading Regge poles \cite{note}. The required Regge cuts are estimated using the absorptive Regge model. This estimate is then used to determine the  sensitivity of  experiments at RHIC to the presence of the Odderon. 
\end{abstract}
\vfill \eject \pagestyle{plain}
\setcounter{page}{1}

\section{Previous work}
Several years ago, Leader and the author \cite{L&T} noticed  that the double spin asymmetry $A_{NN}$ at small $t$ could be a sensitive indicator of the existence of the odderon \cite{history}. This observation was based on earlier work on the use of $A_N$ in the CNI region for proton polarimetry which utilized the enhanced magnitude of the single-spin asymmetry due to the interference of the imaginary part of the strong (nuclear) scattering with the singular Coulomb scattering \cite{buttimore}. Much of that work was devoted to trying to determine the spin dependence of the pomeron coupling which is supposed to be dominant at high energy and is primarily imaginary. We then noticed that the double-spin asymmetry is proportional to the { \it real part} of the interference and that, because the odderon has the opposite $C$ and signature from the pomeron,  its contribution is primarily real and if it exists we might  find an enhanced signal for it in $A_{NN}$. 
It is important to emphasize that we take the odderon to be the $C=-1$ partner of the pomeron and assume that it lies near to $J=1$ at $t=0$.

The five $pp$ amplitudes are defined by

\begin{eqnarray}
\phi_1(s,t) & = & \langle ++|M|++ \rangle , \nonumber \\ \nonumber
\phi_2(s,t) & = & \langle ++ |M|-- \rangle , \\ \nonumber
\phi_3(s,t) & = & \langle +- |M|+- \rangle , \\  \nonumber
\phi_4(s,t) & = & \langle +- |M|-+ \rangle , \\ 
\phi_5(s,t) & = & \langle ++ |M|+- \rangle ,
\end{eqnarray}

and the two asymmetries just mentioned are given by \cite{buttimore}

\begin{eqnarray} \label{eq:asymdef}
A_N \frac{d\sigma}{dt}& =& -\frac{4\pi}{s^2} 
\im \{\phi_5^*(\phi_1 +
\phi_2 +
\phi_3 -\phi_4)\}, \nonumber \\ 
A_{NN} \frac{d\sigma}{dt}& =& 
\frac{4\pi}{s^2}
\{2|\phi_5|^2 + \re
(\phi_1^*
\phi_2 -
\phi_3^* \phi_4) .
\end{eqnarray}
A closely related double spin asymmetry for polarization normal to the beam but {\it in} the place is

\beq
A_{SS} \frac{d\sigma}{dt} =
\frac{4\pi}{s^2}
\{ \re
(\phi_1^*
\phi_2 +
\phi_3^* \phi_4) .
\eeq

Each of these amplitudes, to order $\alpha$ is the sum of an electromagnetic piece $\phi_i^{em}$ and a nuclear piece $\phi_i^{had}$. Although it is not strictly true, for this work for small $|t|$ we will take 
$\phi_1-\phi_3=0$. The normalization is given by
\begin{equation}
\phi_1^{had} + \phi_3^{had}= \f{s}{4 \pi} \sgmtot (i + \rho) \exp{b t/2}
\end{equation}
and $\phi_5$ is assumed to have the form at small $t$
\begin{equation}
\phi_5^{had} = \tau(s) \frac{\sqrt{-t}}{m} \, \phi_1^{had}.
\end{equation}
$\tau(s)$ was determined in \cite{note,firstnote, tojo,Jinnouchi,pjet}.
 We assume throughout that all pieces of the amplitudes have the same slope $b(s)/2 $ in $t$. This is no doubt an oversimplification, {\it faute de mieux}, and when differences in the slopes become known or interesting they can be easly taken into account.

The electromagnetic amplitudes, for small $|t|$ are
\begin{eqnarray}
\phi_1^{em}=\phi_3^{em} = \frac{\alpha s}{t} g^2(t) \nonumber \\
\phi_2^{em}= -\phi_4^{em} = \frac{\alpha s\, \kappa^2}{4\, m^2} g^2(t) \nonumber \\
\phi_5^{em}= -\frac{\alpha s \kappa}{2 m \sqrt{-t}} g^2(t),
\end{eqnarray}
where
\beq
g(t)=\frac{1}{(1 - t/0.71)^2}
\eeq
The Coulomb enhanced $A_N$ contains two pieces: the classic Schwinger piece from the interference between the Coulomb magnetic moment (or spin-flip) piece and the non-flip nuclear piece---this is exactly calculable in terms of measurable quantities, the total cross section and the proton electromagnetic form factor. The other piece  has exactly the same shape in $t$ and but the magnitude depends essentially on Re$(\tau)$. 
In addition there is a purely hadronic piece which vanishes as $t \rightarrow 0$ and so is very small near the CNI peak but grows to comparable values at slightly higher momentum transfer depending mainly on Im$(\tau)$. See RHIC Note \cite{firstnote}. Much of the RHIC spin work of the past few years, experimentally and theoretically, has been directed at determining $\tau$ which is {\it a priori} a function of $s$. 
For $A_{NN}$ the story is similar, but different in essential ways. The electromagnetic part of $\phi_2$ is not singular at $t=0$ so only the term $\re({\phi_1^{em\, *} \phi_2}^{had})$ will be enhanced and it will be proportional to $1/t$ rather than $1/\sqrt{t}$. The enhancement is proportional to the completely unknown ratio of $\phi_2^{had}/\phi_1^{had}$. In order to illustrate the shape of the signal in the next figure,we used 5\% magnitude at $t=0$ for either the pomeron or the odderon. The characteristic difference is striking.  (The magnitude seemed an interesting choice because it would be observable in $pp2pp$. \cite{pp2pp}) We have assumed here that the energy is sufficiently high that only the pomeron and odderon contribute and we assumed they were both simple poles at $J=1$ when $t=0$. See Fig.1. \cite{L&T}
\begin{figure}[htb]
\begin{center}
\centerline{\epsfxsize=4.1in\epsfbox{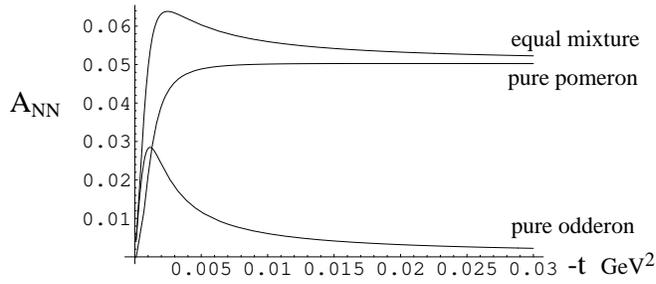} } 
\caption{{\bf Not a prediction: for motivation only.}}
\end{center}
\end{figure}

This is meant as an illustration, an idea, and we will see that there are significant modifications to this. Now, as the result of analysis of the data from the $p$-carbon and $p$-jet experiments at the AGS and at RHIC, we have some knowledge of the the spin dependence of the important Regge couplings. We can use these to make some estimates ---model dependent, to be sure---of the magnitude and energy dependence of $A_{NN}$. 

\section{Regge pole couplings}

Recent work directed at determining the energy dependence of $A_N$ for proton polarimetry is described in \cite{firstnote} and in \cite{note}. The only practical framework for studying energy dependence of small $t$ processes is the Regge theory. It has been reasonably successful over the years, especially in describing unpolarized amplitudes. The principal drawback is that there is no way to reliably calculate the couplings of the Regge poles to the particles in question; these are determined from experiment and then used at other energies or in other processes.

For $pp$ scattering, for each Regge pole we will need two couplings, or ``residues", non-flip $\beta_{nf}^R$ and flip $\beta_{f}^R$.  It is a basic feature of the theory that Regge poles factorize at their two vertices so if the non-flip amplitude $\phi_1^{had}$ is given by
\beq
\phi_1(s,t)=-\frac{1}{8 \pi}\Sigma_R \beta_{nf}^R(t) \beta_{nf}^R(t) (s/s_0)^{\alpha_
R}\f{ (\exp{(-i \pi \alpha_R)} +\mathcal{S}_R)}{\sin{\pi \alpha_R}}.
\label{nonflip}
\eeq
where $\mathcal{S}_R$ is the signature of the Regge pole $R$, then for the single flip amplitude $\phi_5$, $\beta_{nf}^R(t) \beta_{nf}^R(t) \rightarrow  \beta_{nf}^R(t) \beta_{f}^R(t)$ and for the double-flip amplitude $\phi_2$, $\beta_{nf}^R(t) \beta_{nf}^R(t) \rightarrow  \beta_{f}^R(t) \beta_{f}^R(t)$. The residues are all real in the physical scattering region $t \leq 0$ and therefore the phase of each term is determined through the energy dependence and the signature factor. \cite{berger, I&W}.

Several groups have used the form Eq.(\ref{nonflip}) to determine best values for the various trajectory fucntions $\alpha_R$ and the corresponding residues for non-flip amplitudes. Reasonable fits are found in a variety of different models including multiple Regge pole and Regge cuts. The work of \cite{note} is built upon the simple Regge pole model of Cudell et al \cite{Cudell}. The model consists of three poles: (1) the pomeron, (2) a $C=-1$ Regge pole, a composite of $\rho$ and $\omega$, and (3) a $C=+1$ Regge pole, a composite of $f$ and $a2$. The parametrization seems to be adequate for $s \geq 45\, \rm{GeV}^2$, where we will make use of it.

The spin flip dependence is introduced by way of
\beq
\beta_f^R(t) = \tau^R\f{ \sqrt{-t}}{m}\,\beta_{nf}^R(t)
\eeq
The factor $\sqrt{-t}/m$ is required in the spin-flip amplitudes by angular momentum conservation. $\tau^R$ is real and independent of $s$ but in principle can have additional $t$-dependence;l we will neglect it over the small CNI region. Correspondingly, we use the parametrization
\beq
\phi_5(s,t) = \tau(s)\f{ \sqrt{-t}}{m}\, \phi_1(s,t).
\eeq
Note that $\tau(s)$, in contrast to the $\tau^R$ is in general complex and energy dependent. At any given energy, both the real and imaginary parts can be determined by analysis of $A_N(t)$ over the CNI region. This can be done, using the proton-carbon data for the $I=0$ exchanges (cf. \cite{K&T}) at the AGS 24 GeV/c and at RHIC 24 and 100 GeV/c \cite{firstnote,tojo, Jinnouchi, pjet} and the proton-jet data at 100 GeV/c  for the $I=1$ exchanges \cite{pjet}. The beam polarization must be known at one energy for each target, but that is sufficient to determine the three spin-flip ratios we need.

The values  found are tentative and will be revised as soon as more data is made available. The resulting spin-flip factors used here are

\beqn
\tau_P=0.09 \pm .015 \\ \nonumber
\tau_{+}=-0.32 \pm .08\\ \nonumber
\tau_{-}=1.06 \pm 0.64
\eeqn
These values and the procedure used to obtain them are reported in \cite{note}; a complete report is under preparation.
Note the much larger size of the negative $C$ spin-flip, mostly due to the $\rho$. Also notice the very large error on it.  The old analysis by Berger et al \cite{berger} of the much lower energy data from the Argonne ZGS also gives an exceptionally large value for the $\beta_f^\rho$, similar to the value found here.

There is no sign of the odderon in the small-$t$ unpolarized scattering \cite{block}  so we will take its non-flip coupling to be zero, and its spin-flip coupling to be $\sqrt{-t/m^2} \, \beta_f^O$.  This leads to a double-flip amplitude of the form,
\beqn
\phi_2^O= -\frac{t}{8 \pi m^2} \, \beta_f^{O \, 2} \, \frac{(1- e^{-i \pi \alpha_O})}{\sin{\pi \alpha_O}}(s/s_0)^{\alpha_ O} e^{b(s) t/2}
\label{oddpole}.
\eeqn
Following Nicolescu \cite{Nicolescu} we take $\alpha_O =0.96$ and neglect its $t$-variation  over the very small CNI region. A simple Regge pole analysis of $\phi_2$ will necessarily lead to no CNI enhancement for $A_{NN}$, odderon or no odderon: because of factorization of the pole couplings, $\phi_2$ must vanish the same way that $\phi_4$ does as $t \rightarrow 0$; i.e they both vanish proportional to $t$. This cancels the singularity in the Coulomb ampitude and leads to no enhancement at all. This is shown in Fig. 2 for both no odderon coupling and for strong odderon coupling, with the odderon flip equal to the $\rho$ flip. (Note that the first is multiplied by 100 to put it on the same figure.)
 \begin{figure}[htb]
\begin{center}
\centerline{\epsfxsize=4.1in\epsfbox{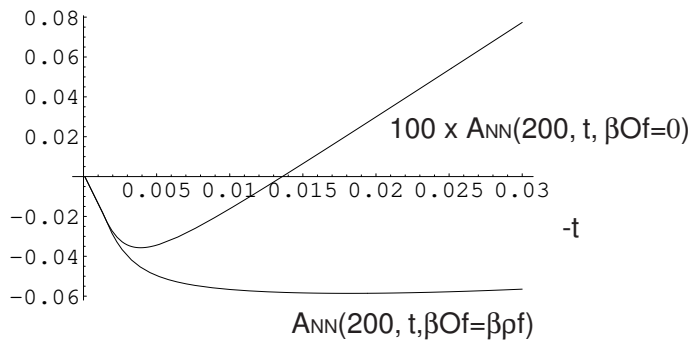} } 
\caption{{\bf Regge pole model prediction for $A_{NN}$ at $s=200$}}
\label{poles}
\end{center}
\end{figure}
Because factorization leads to suppression of the Regge pole contribution to $\phi_2$  by a factor $t$, it is natural to examine the cut contribution generated by these poles.

\section{Regge cuts and $A_{NN}$}  

For a long time Regge cuts have been known to be present in any Regge theory ; for example, they result from iterated exchange of different Regge poles. They are subtle and difficult to calculate confidently because there are important, sometimes total, cancellations between graphs that look very different. 
Regge cuts have some important properties: (1) their energy dependence is $$s^{\alpha_P(0) +\alpha_R(0) -1}/\log{s},$$ (2) they have the same charge conjugation and signature as the corresponding pole; (3) they do not have a definite parity \cite{l&j}. To see the importance of this one can examine the table extracted from \cite{buttimore}.
\begin{table}[h]
\centering
$\begin{array}{|c|c|c|}
\hline
\mbox{Class 1} & \mbox{Class 2} & \mbox{Class 3} \\ 
\tau = P = C	& \tau = -P = -C  &
\tau = -P = C
\rule[-.5cm]{0cm}{10mm} \\ \hline \phi_+,\phi_5, \phi_2 \!-\! \phi_4 &
\phi_- &
\phi_2 + \phi_4 \rule[-.5cm]{0cm}{10mm}\\ \hline I\hspace{-1.4mm}P, O,
\rho, \omega,  f, a_2 & a_1 &
\pi, \eta, b
\rule[-.5cm]{0cm}{10mm}\\ \hline 
\end{array}$
\caption{\sl Classification of $pp$ amplitudes by exchange symmetries 
and the
associated Regge poles.}
\end{table}

Notice that if the Regge singularity has a definite parity, then $\phi_2$ would necessarily vanish at $t=0$ because $\phi_4$ does, independent of factorization. However, if the cut has a mixed parity, then by combining the Class 1 and Class 3 amplitudes, $\phi_2$ and $\phi_4$ can behave independently. We will happily find that the explicit calculation of the cut verifies this property.

I don't propose to attempt a rigorous calculation of the cuts associated with the Regge poles; rather I will use a simple, intuitive model that has been long used and hope that the result is reasonable.  It is called the absorptive Regge model \cite{frank}; basically it amounts to a double-scattering on-shell with one scattering given by the Regge exchange and the other by pomeron exchange. For the cut generated by the pomeron and the above defined odderon this yields
\beqn
\phi_2^Ocut=-\frac{ \sigma_{\rm tot}}{2 \pi b(s) } (1 - i \rho) \beta_f^{O 2} (s/s_0)^{\alpha_O } \f{1}{8 \pi b(s) m^2}\, \f{(1-e^{-i \pi \alpha_0})}{sin(\pi \alpha_0)} e^{b(s)\, t/4},
\eeqn
again assuming that the odderon non-flip coupling is negligible. Notice that the factor $t$ seen in Eq.(\ref{oddpole}) is here replaced by $1/b$, the slope of the amplitudes.  Strictly speaking this is the pomeron-odderon cut only asymptotically since $\sigma_{\rm tot} $ contains contributions from the other Regge poles in addition to the pomeron; this is the full absorptive Regge model. Also it should be noted that in \cite{frank} certain fudge factors are included to account for inelastic in addition to elastic absorption. We disregard these factors here.

The same calculation leads to a cut contribution to $\phi_4$.  As required by angular momentum it vanishes as $t\rightarrow 0$ but is otherwise similar to $\phi_2$:
\beqn
\phi_4^Ocut=- \frac{b(s) t}{4}\, \phi_2^Ocut .
\eeqn 
Evidently it is small compared to $\phi_2$ throughout the CNI region.

\begin{figure}[!]
\begin{center}
\centerline{\epsfxsize=4.1in\epsfbox{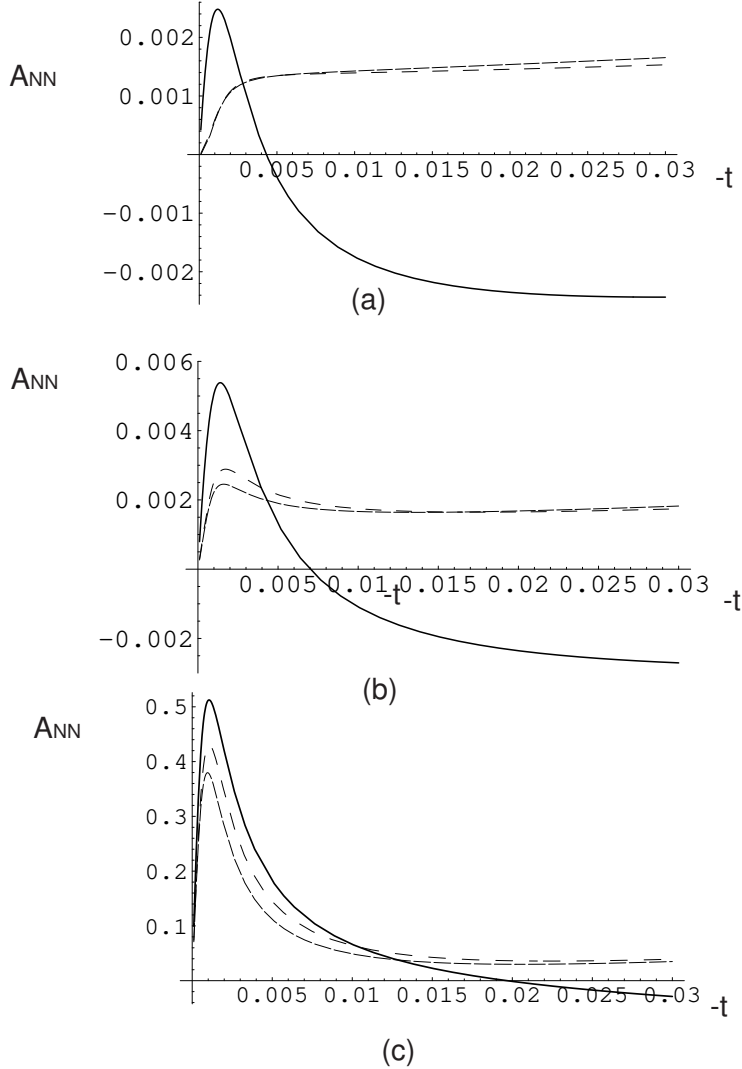} } 
\caption{{\bf for $s=200$ (solid), $200^2 $ (dash space), $s=500^2$ (dash short-space) with odderon couplings: (a) no odderon, (b) equal to Pomeron flip coupling, (c) equal to $\rho$ flip coupling}}
\label{cuts}
\end{center}
\end{figure}

We are principally interested in the odderon-pomeron cut here so we need to make some guesses at  the totally unknown flip couplings of the odderon.  In the next figure we illustrate cases of $A_{NN}$ at $s=200, s=200^2, s=500^2$ for a range of couplings: (a) no Odderon  $\beta_f^O=0$, (b) weak Odderon  $\beta_f^O=\beta_f^P$ and (c) strong Odderon  $\beta_f^O = \beta_f^{\rho}$.  %

The first plot is the prediction of the simple Regge model with cuts based on the flip coupling measured with $A_N$. It is disappointingly small.  Even with an odderon with pomeron size coupling included, the signal is tiny. Only with the large $I=1, C=-1$ coupling does the odderon give a strong signal. (The curves (a) and (b) at the lowest energy show enhancement peaks because the phases assumed in Fig.~1 are not correct at such low energy.)

Fig.~4 shows the energy dependence of $A_{NN}$ at $t=-.001$, near the peak of the curve, as a function of energy for the three standard spin-flip possibilities. In order to put them all on the same figure we have reduced the $\beta_{ f}^{\rho}$ curve by a factor of 50.
\begin{figure}[htb]
\begin{center}
\centerline{\epsfxsize=4.1in\epsfbox{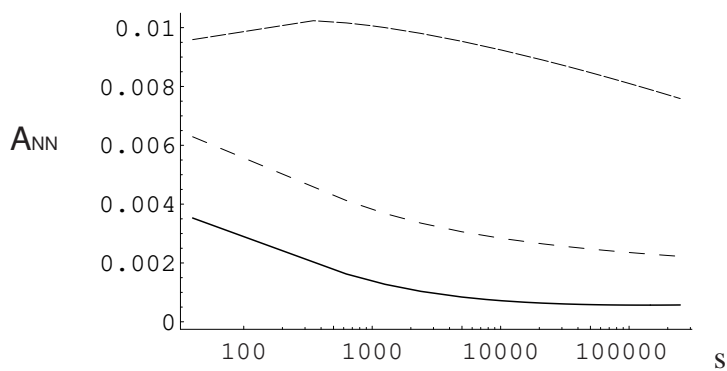} } 
\caption{{\bf  $A_{NN}$ near the peak as a function of $s$ for $\beta_f^O=0$ (solid),  $\beta_f^O=\beta_f^P $  (large gaps) and (scaled by a factor of 1/50) $\beta_f^O=\beta_f^\rho$ (small gaps)}}
\label{cuts}
\end{center}
\end{figure}
It is seen that there is a generally slow decrease in the asymmetry as the energy moves higher.

 It is interesting to look also at $A_{SS}$ for comparison to $A_{NN}$; that might be measured in the same experiment, or a linear combination of $A_{SS}$ and  $A_{NN}$ depending on the relative orientation of the polarization and the detectors might be measured. The predictions are shown in Fig.~5 corresponding to the same variable values as in Fig.3.
 \newpage
\begin{figure}[!]
\begin{center}
\centerline{\epsfxsize=4.1in\epsfbox{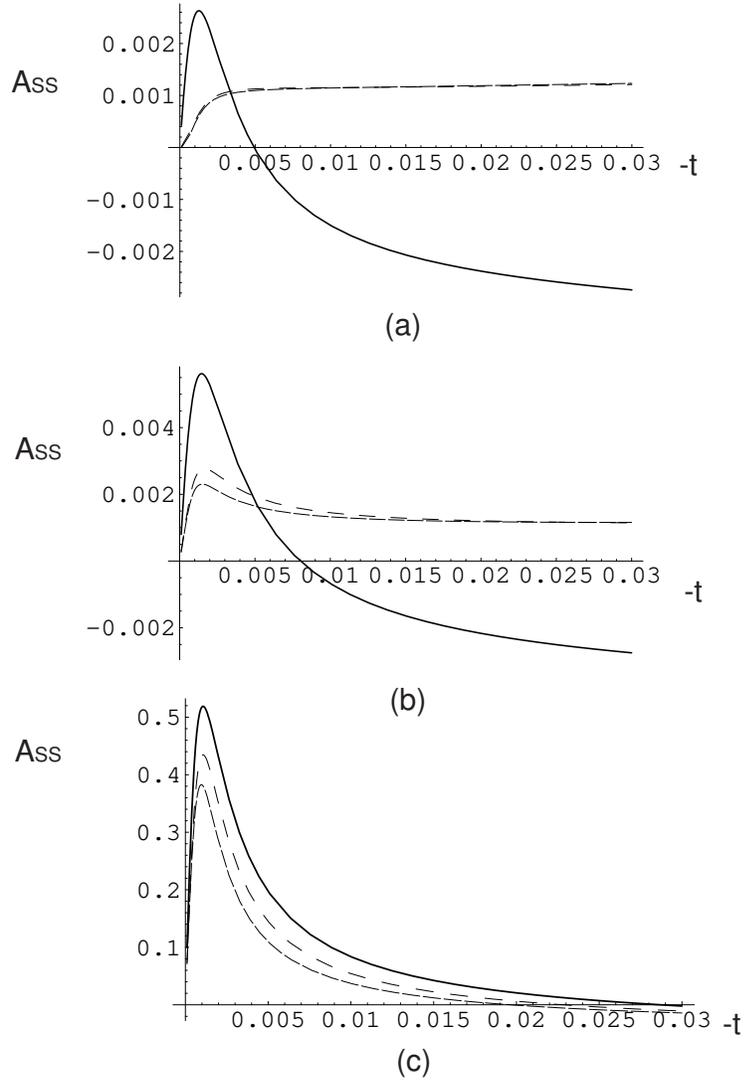} } 
\caption{{\bf same as Fig. ~3}}
\label{cuts}
\end{center}
\end{figure}
The curves are not identical but are very similar to those for  $A_{NN}$, and the measurement should not be very sensitive to the polarization orientation.
\section {Conclusions} 
We have here attempted to use new information about the spin dependence of $pp$ elastic scattering to investigate the idea that this could be a good tool to observe the presence of an Odderon if it exists \cite{L&T}. This is intended as a complement to the several other experimental signatures of the Odderon  that have been suggested \cite{collective}. For a summary see the workshop proceedings and especially the summary talk by C. Ewarz \cite{workshop}. 

We have found that in the absence of an Odderon the double-spin asymmetry $A_{NN}$ is very small throughout the CNI region, of order 0.002 or smaller. A small Odderon coupling of the order of the Pomeron spin-flip coupling is not much larger but has a very different energy dependence and could in principle be distinguished from the former case. A large Odderon coupling of order the $\rho$ spin-flip coupling gives an enormous effect with the characteristic shape proposed in \cite{L&T}.

These conclusion must be considered preliminary because much of the experimental work the calculations are based on is still preliminary and some necessary theoretical assumptions (e.g. the formula for cuts) have obvious weaknesses and might have to be changed. It does seem however that some very interesting physics could be learned from the study of $A_{NN}$ in the CNI region of $pp$ elastic  scattering.

{\bf Acknowledgement}
Thanks to Wlodek Guryn for his strong and constant encouragement  for this work.

\end{document}